\documentclass[10pt,a4paper,oneside,english]{elsart}
\usepackage[T1]{fontenc}
\usepackage[latin1]{inputenc}
\usepackage{float}
\usepackage{graphicx}
\usepackage{amssymb}
\usepackage{url}

\usepackage{babel}
\usepackage{epsfig}
\makeatother
\begin{document}
\begin{frontmatter}

\title{Double power laws in income and wealth distributions}

\author{Ricardo Coelho},
\ead{rjcoelho@tcd.ie}
\author{Peter Richmond},
\author{Joseph Barry}
\author{\and Stefan Hutzler}

\address{School of Physics, Trinity College Dublin, Dublin 2, Ireland}

\begin{abstract}
Close examination of wealth distributions reveal the existence of two distinct power law regimes. The Pareto exponents of the super-rich, identified for example in rich lists such as provided by Forbes are smaller than the Pareto exponents obtained for top earners in income data sets. Our extension of the Slanina model of wealth is able to reproduce these double power law features.
\end{abstract}

\end{frontmatter}

\section{Introduction}

The first person to study the topic of wealth distributions in a quantitative manner, Pareto, was trained as an engineer \cite{Pareto_1896}. In recent years, it is the physics community who have made significant contributions to the topic, again by focusing not only on theoretical methodologies \cite{{Malcai_PRE66_031102_2002},{Chatterjee_PhysA335_155_2004},{Repetowicz_PhysA356_641_2005},{Coelho_PhysA353_515_2005},{Richmond_PhysA370_43_2006}} but also making comparisons of their results with empirical data \cite{{Souma_Fractals9_463_2001},{Dragulescu_EPJB20_585_2001},{Clementi_PhysA350_427_2005},{Sinha_PhysA359_555_2006},{Repetowicz_article}}. For a recent detailed review of the subject see \cite{Indian_Book}.

What does seem clear from the mounting evidence is that income and wealth distributions across societies everywhere follow a robust pattern and that the application of ideas from statistical physics can provide understanding that complements the observed data. The distribution rises from very low values for low income earners to a peak before falling away again at high incomes. For very high incomes it takes the form of a power law as first noted by Pareto. The distribution is certainly not uniform. Many people are poor and few are rich.

The cumulative probability corresponds to the probability of finding earners that have an income bigger or equal to a certain amount of income, $m$. For values of $m$ less than the average income it decreases slowly from its maximum value $1$. For values roughly high than the average it follows a power law: 
\begin{equation}
P(_{>}m)=m^{-\alpha}
\end{equation}
where $\alpha$ is the Pareto exponent. 

\begin{figure}[H]
  \begin{center}
    \epsfysize=80mm
    \epsffile{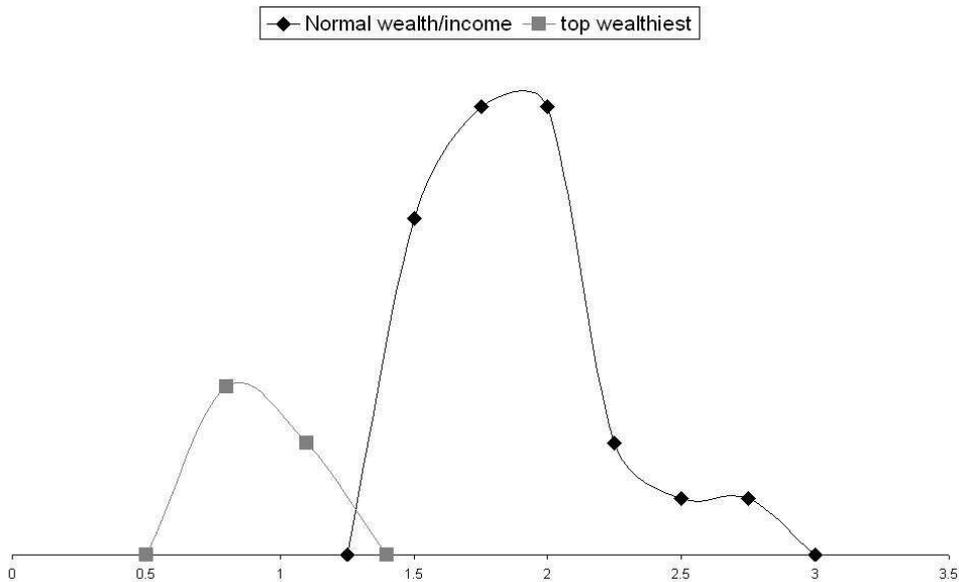}
    \caption{Distribution of the Pareto exponents found by different authors in the last decade. The black curve is from data sets taken from tax/income databases. The grey curve is from super-rich lists, such as Forbes. The Pareto exponent for the top richest is around $1$ while for the ``normal'' rich people is around $2$ (data taken from Table 5.2 in reference \cite{Indian_Book}).}
    \label{figure_1}
  \end{center}
\end{figure}

\begin{figure}[H]
  \begin{center}
    \epsfysize=90mm
    \epsffile{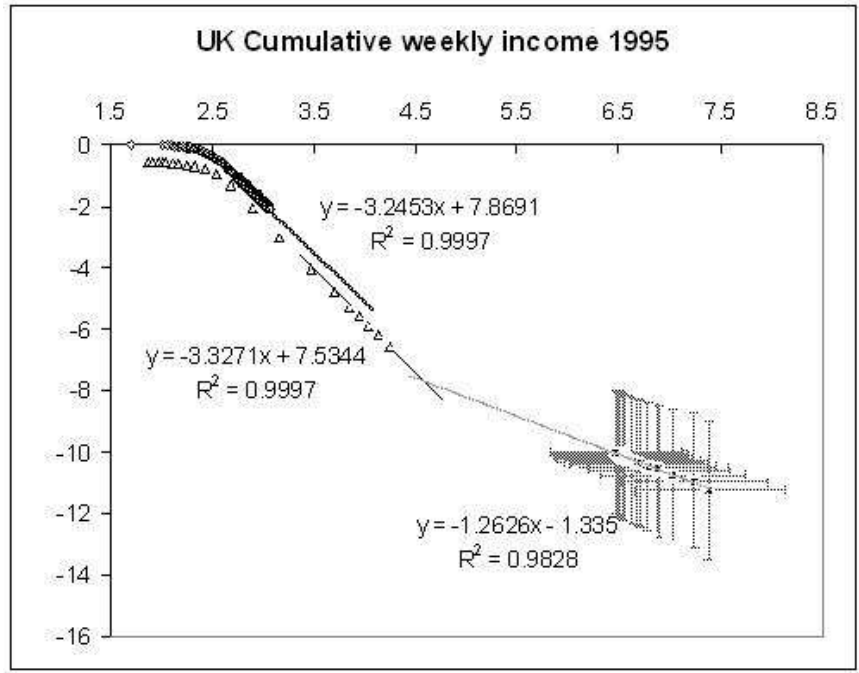}
    \caption{Distribution of the cumulative weekly income in U.K. for $1995$. The left side curves represent income for $1995$ from two different sources and a similar Pareto exponent is achieved for the high end of these curves, $\sim 3.2 - 3.3$. The right side curve represent an estimation of the income, in $1995$, for the top richest in U.K. In this case the Pareto exponent is lower and around $1.3$.}
    \label{figure_2}
  \end{center}
\end{figure}

Looking closely at results for income and wealth distributions around the world (Table 5.2 in reference \cite{Indian_Book}) we see that the values for the exponents for wealth/income data sets, and data that concerns only the top wealthiest people in society differ. Figure \ref{figure_1} shows the distribution of the Pareto exponents when we take these different origins of the data into account. The average Pareto exponent is approximately $2.0$ for the top earners in tax/inheritance statistics, and just below $1.0$ for the super-rich.
We believe that the studies of wealth that are based on tax/income generally do not include the wealth of very rich people. A further indication of two power law regimes is the study of Souma \cite{Souma_Fractals9_463_2001}. In Figure $1$ of his paper \cite{Souma_Fractals9_463_2001}, Souma found a Pareto exponent of $2.06$ in the high end. However, we see an indication of a second power law for the top richest (higher than $3000$ million yen) which we estimate as an exponent below $1.0$ based on his figure.
Yet a further indication of two power laws comes from our analysis of U.K. data. In Figure \ref{figure_2} we show data for the cumulative distribution of incomes in the UK for the year $1995$. The upper curve is calculated from survey data and tends to a power law which was confirmed by data obtained by Cranshaw \cite{Cranshaw} from the UK Revenue Commissioners. The lower curve is calculated using the U.K. New Income Survey data, which takes a $1\%$ sample of all employees in Great-Britain. The slight shift in the two curves is due to uncertainty in a normalisation factor but the power law is clearly seen and extends from weekly incomes of just under $\pounds 1000$ per week up to around $\pounds 30000$ per week. Over this region the Pareto exponent is $\sim 3.3$. This might be assumed to be the end of the story with the power law being associated with Pareto's law. However from data published by Sunday Times \cite{SundayTimes} for the wealth of billionaires in U.K. for $2006$, we can make an estimate of the income in $1995$ generated by the wealth. In order to move from $2006$ back to $1995$, we made some creative estimations. First, we said that probably the wealth of the top richest group had increased in proportion to the stock market over the period $1995$ to $2006$. This index has roughly doubled in that time, so in $1995$, the wealth is roughly $50\%$ of the $2006$ value. Then we assumed this wealth generated an income from being invested and the interest rate was around $4\%$ per annum. This yields a second power law with Pareto exponent $\sim 1.26$. This suggests what many people believe to be true, namely that the super wealthy pay less tax as a proportion of their income than the majority of earners in society!

\section{Wealth models}
A number of models have been proposed to account for the distributions of wealth in society. One class that might be considered to constitute a {\it mesoscopic} approach is based on a generalised Lotka Volterra models \cite{{Malcai_PRE66_031102_2002},{Solomon_IJMP12_333_2001},{Solomon_EPJB27_257_2002}}. Other {\it microscopic} models invoke  agents that exchange money via pairwise transactions according to a specific exchange rule. The results from these latter models depend critically on the nature of the exchange process. Two quite different exchange processes have been postulated. The first by Chakraborti and colleagues \cite{{Chakraborti_EPJB17_167_2000},{Patriarca_PRE70_016104_2004}} conserves money during the exchange and allows savings that can be a fixed and equal percentage of the initial money held by each agent. This yields the Boltzmann distribution. Allowing the saving percentage to take on a random character then introduces a power law character to the distribution for high incomes. The value of the power law exponent however can be shown to be exactly one \cite{Repetowicz_PhysA356_641_2005}. Only a slight variation of the exponent is achieved by attributing memory to the agents \cite{Repetowicz_article}.

On the other hand, the model of Slanina \cite{Slanina_PRE69_046102_2004} assumes a different exchange rule. It also allows creation of money during each exchange process and the solution is not stationary. One must normalise the amount of money held by an agent with the mean value of money within the system at time $t$. In this way a stationary solution for the distribution of the normalised money can be obtained. Such a procedure must also be applied to obtain a solution from the Lotka Volterra approach and it is interesting to see that the final results for both methods yield distribution functions of the same form. Detailed numerical comparisons with the data suggest that this form gives a good fit to the data below the super rich region \cite{Indian_Book}.

\section{Expansion of Slanina's model}

\begin{figure}[H]
  \begin{center}
    \epsfysize=100mm
    \epsffile{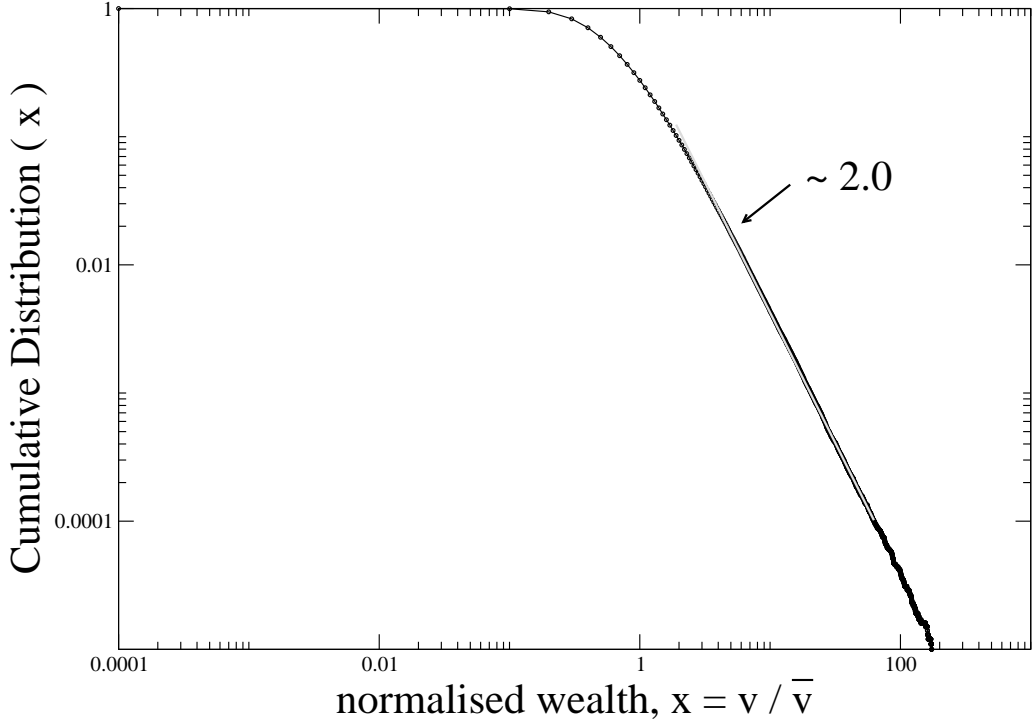}
    \caption{Cumulative distribution of wealth in a simple Slanina model, for $10^4$ agents trading $10^3 \times N $ times and averaged over $10^3$ realisations. The percentage of wealth exchanged ($\beta$) is equal $0.005$ and the percentage of wealth injected in the system ($\epsilon$) is $0.1$. The Pareto exponent for the higher end is $\sim 2.0$.}
    \label{figure_3}
  \end{center}
\end{figure}

Slanina's model \cite{Slanina_PRE69_046102_2004} involves the pairwise interaction of agents, which at every exchange process also receive some money from outside. The time evolution of trades is represented as:
\begin{equation}
\left( \begin{array}{c}
v_i(t+1) \\
v_j(t+1)
\end{array}
\right) = \left( \begin{array}{cc}
1 + \epsilon - \beta & \beta \\
\beta & 1 + \epsilon - \beta
\end{array}
\right)
\left( \begin{array}{c}
v_i(t) \\
v_j(t)
\end{array}
\right)
\end{equation}
where $v_i(t)$ is the wealth of agent $i$ at time $t$ ($i=1,\dots,N$, where $N$ is the total number of agents), $\beta$ quantifies the percentage of wealth exchanged between agents and $\epsilon$ measures the quantity of wealth injected in the system at every exchange. In the simplest case, the values of $\beta$ and $\epsilon$ are kept constant for all trades. This results in a power law for the rich end at the normalised distribution of wealth, i.e. the distribution of $x_i(t)=v_i(t)/\bar{v}(t)$ where $\bar{v}(t)$ is the mean wealth at time $t$ ($\bar{v}(t) = \sum_{i=1}^N v_i(t)/N$). An approximation of the Pareto exponent is given by Slanina paper \cite{Slanina_PRE69_046102_2004} as:
\begin{equation}
\alpha \sim \frac{2 \beta}{\epsilon^2} + 1
\label{Slanina_exponent}
\end{equation}
apart from some correction in the $\epsilon$ term.
To check the accuracy of this approximation, we ran some simulations for $10^4$ agents trading $10^3 \times N $ times and averaged over $10^3$ realisations. The percentage of wealth exchanged ($\beta$) was set to $0.005$ and the percentage of wealth injected in the system ($\epsilon$) to $0.1$. Fitting a power law to the high end of our distribution in Figure \ref{figure_3}, we find an exponent of $2.0$ in excellent agreement with the value of $2.0$ of equation \ref{Slanina_exponent}

Our expansion of Slanina's model is given by making $\beta$ a function of $v$, $\beta(v)$. The main conclusion that we can take from this wealth dynamic is that a double power law arrives from the difference between the percentage of money that agents put in the society for trade. This difference can be related with different levels of fear to risk or from some economical issues related with taxation. This results in the following update rule:
\begin{equation}
\left( \begin{array}{c}
v_i(t+1) \\
v_j(t+1)
\end{array}
\right) = \left( \begin{array}{cc}
1 + \epsilon - \beta(v_i) & \beta(v_j) \\
\beta(v_i) & 1 + \epsilon - \beta(v_j)
\end{array}
\right)
\left( \begin{array}{c}
v_i(t) \\
v_j(t)
\end{array}
\right).
\end{equation}
Here we consider the simplest case, i.e.:
\begin{equation}
\begin{array}{c}
\beta(v)
\end{array}
 = \left\{ \begin{array}{cc}
\beta_1, & v < n \bar{v}(t) \\
\beta_2, & v \geq n \bar{v}(t) 
\end{array}
\right.
\begin{array}{c}
, \beta_1 > \beta_2
\end{array}
\end{equation}
If an agent has wealth higher than a threshold ($n$ times the average wealth, $\bar{v}(t)$), the second parameter ($\beta_2$) will be used. The threshold adopted in these simulations is $10 \bar{v}(t)$, so if agent $i$ or $j$ have a mean wealth higher than this threshold they will trade a different percentage as if they would have a smaller amount.
 
\begin{figure}[H]
  \begin{center}
    \epsfysize=100mm
    \epsffile{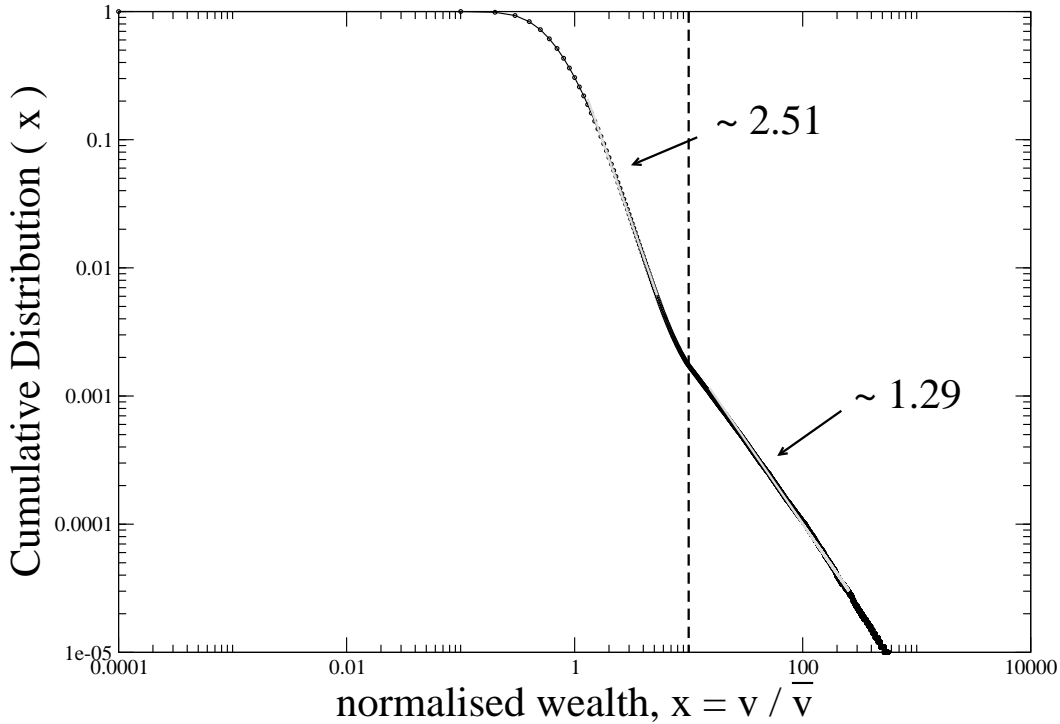}
    \caption{Cumulative distribution of wealth in expanded Slanina model. The values for number of agents, time steps, realisations and percentage of wealth injected in the system ($\epsilon$) are the same as used in Figure \ref{figure_3}. The percentage of wealth exchanged ($\beta$) if the agent has wealth smaller than $10 \bar{v}(t)$ is $0.01$ $0.00125$ and if the agent has wealth higher or equal to $10 \bar{v}(t)$ is $0.00125$. Two different Pareto exponents appear in different parts of the distribution. One for what we call rich people is around $\sim 2.5$ and a second one for the top richest is around $\sim 1.3$. The vertical dashed line shows the threshold that we choose for different $\beta$'s values.}
    \label{figure_4}
  \end{center}
\end{figure}

To simulate a society like the U.K., where two Pareto exponents exist, one for the top earners around $3.0$ and another one for the super-rich around $1.5$, we have chosen the parameters $\beta$ and $\alpha$ according to equation \ref{Slanina_exponent}, i.e. $\beta_1=0.01$, $\beta_2=0.00125$ and $\epsilon=0.1$. Figure \ref{figure_4} shows the result of our simulations. Two distinct power laws are visible, one in the regime between $\bar{v}(t)$ and $10 \bar{v}(t)$ and another one for wealth larger than $10 \bar{v}(t)$. The Pareto exponents are $2.51$ and $1.29$, respectively, and thus differ from the prediction of equation \ref{Slanina_exponent}. However in our case, this prediction should only be taken as a first order approximation, since we are essentially dealing with two societies (each specified by its respective $\beta$ values) which are interacting. Agents switch between their interaction parameters according to their relative wealth.

The main success of the modified Slanina model is thus the reproduction of two power laws regimes. Contrary to this, and to some surprise, calculations based on a similarly modified Lotka Volterra model did not result in such double power law distribution.
A better accuracy of the theoretical results should be achieved in future work, where we intend to find the solution for the case of two Pareto exponents in the same wealth distribution.

\section{Conclusion}

As was discussed in \cite{Indian_Book}, progress in understanding the details of wealth distribution is invariably linked to obtaining data sets that encompass the entire population of a country. It appears that at present, this information is only available for a few countries, for example Japan (Souma \cite{Souma_Fractals9_463_2001}). Generally, the super-rich are not included in income data. Published wealth lists are estimates, but for the moment might well remain the only public source for the information on these top earners. We hope that analyses of the kind we have made in this paper encourages the release of more detailed income data over the entire income range. Only with more complete data sets will we be able to properly understand these complex economic systems.

\begin{ack}
R. Coelho acknowledges the support of the FCT/Portugal through the grant SFRH/BD/$27801/2006$. J. Barry is funded by IRCSET (Irish Research Council for Science, Engineering and Technology). The authors also acknowledge the help of COST (European Cooperation in the Field of Scientific and Technical research) Action P10. 
\end{ack}

\end{document}